\documentstyle[epsf]{mn}

\newcommand{\etal}{{\rm et al.~}}
\newcommand{\Mpc}{$h^{-1}$~{\rm Mpc}}

\title[Correlation function]{The supercluster--void network III.\\
       The correlation function as a geometrical statistic}

\author[J. Einasto et al.]
{J. Einasto$^1$,  M. Einasto$^1$, P. Frisch$^2$, 
S. Gottl\"ober$^3$, V. M\"uller$^3$, V. Saar$^1$, 
\newauthor 
A.A. Starobinsky$^4$, D. Tucker$^{3,5}$ \\
$^1$ Tartu Observatory, EE-2444 T\~oravere,
Estonia \\
$^2$ G\"ottingen Univ. Observatory, Geismarlandstr. 11,
D-37083 G\"ottingen, Germany \\
$^3$ Astrophysical Institute Potsdam, An der Sternwarte 16,
D-14482 Potsdam, Germany \\
$^4$ Landau Institute Theoretical Physics, Moscow, Russia \\
$^5$ Fermilab, Batavia, IL 60510, USA}

\date{Accepted:  1997;  Received  1996}
\pubyear{1997}

\begin{document}

\maketitle

\begin{abstract} 

We investigate properties of the correlation function of clusters of
galaxies using geometrical models. We show that the correlation
function contains useful information on the geometry  of
the distribution of clusters. On small scales the correlation function
depends on the shape and the size of superclusters.  On large scales
it describes the geometry of the distribution of superclusters.  If
superclusters are distributed randomly then the correlation function
on large scales is featureless.  If superclusters have a quasi-regular
distribution then this regularity can be detected and measured by the
correlation function.  Superclusters of galaxies separated by large
voids produce a correlation function with a minimum which corresponds
to the mean separation between centres of superclusters and voids,
followed by a secondary maximum corresponding to the distance between
superclusters across voids. If superclusters and voids have a tendency
to form a regular lattice then the correlation function on large
scales has quasi-regularly spaced maxima and minima of decaying
amplitude; i.e., it is oscillating. The period of oscillations is
equal to the step size of the grid of the lattice.

We also calculate the power spectrum and the void diameter distribution
for our models and compare the geometrical information of the correlation
function with other statistics. We find that geometric properties (the
regularity of the distribution of clusters on large scales) are better
quantified by the correlation function. 

We also analyse errors in the correlation function and the power
spectrum by generating random realizations of models and finding the
scatter of these realizations.

\end{abstract}


\section{INTRODUCTION}
 
The correlation function is one of the most frequently used statistics
in cosmology. Conventionally the correlation function and the power
spectrum are considered together to characterise the basic clustering
properties of  galaxies or of clusters of galaxies. They complement
each other in the sense that the power spectrum describes the structure
in  Fourier space and the correlation function in real space.

Intuitively, it is clear that, on small scales comparable to
the size of systems of galaxies, the galaxy correlation function should
characterise the distribution of galaxies in systems of galaxies. On
these scales the galaxy correlation function is large  -- this
is the manifestation of the clustering of galaxies.  With increasing
scale  the clustering gets weaker and at some scale the
correlation function approaches zero.  Traditionally the correlation
function is expressed in log--log units and is calculated only up to the
zero crossing separation.  This  separation is sometimes used as a
parameter characterising the transition from clustering to homogeneity.

The correlation function of clusters of galaxies has a similar
behaviour, only it is shifted towards larger scales. The clustering
scale is here about a factor of 4 times larger than that for galaxies. 

On large scales the correlation function of galaxies depends primarily
on the distribution of systems of galaxies. Thus, properties of the
distribution of these systems influence the behaviour of the
correlation function on large scales.  Similarly, the correlation
function of clusters of galaxies on large scales depends on the
distribution of systems of clusters, i.e. superclusters of galaxies.
Einasto \etal (1997a,b, hereafter E97 and Paper II)
determined the correlation function of  clusters belonging to rich
superclusters using data in a distance interval $\sim 650$ \Mpc. The
main finding of this study was the detection of a series of
almost regularly spaced maxima and minima. This property of
the correlation function was called {\it oscillations}. The period of
oscillations was found to be $115\pm 15$~\Mpc.

This behaviour of the correlation function emphasises the presence of
certain regularities in the distribution of clusters of galaxies. A
regular network of clusters of approximately the same scale was
actually noticed  by  Tully \etal (1992) and Einasto \etal (1994,
1997c, hereafter Paper I).

These considerations have motivated us to perform an analysis of the
correlation function as a geometrical statistic.  Such analysis helps us to
understand the relation between the geometry of the distribution of clusters
and superclusters and the correlation function. We shall use simple
geometrical models of the particle distribution and calculate the correlation
function for these models with the aim to find the relation between the
correlation function and the geometry of the cluster distribution.  Geometric
information contained in the correlation function has been studied so far only
for relatively small separations. Bahcall, Henriksen, \& Smith (1989) used a
model in which galaxies were located on the surfaces of spherical shells and
clusters of galaxies resided at the intersections of the shells.  Several
models are discussed by Einasto (1992) with the aim of giving a geometrical
interpretation for the galaxy correlation function on small scales.  It was
also noted that if the systems of galaxies are located in a regular grid then
the correlation function has a secondary maximum at a scale which corresponds
to the grid length (step).

In contrast to previous studies we focus our attention in the present
paper on large scales.  On these scales the errors of the correlation
function are of the same order as the function itself, thus a careful
error analysis is crucial.  Additionally we found for our models the
power spectrum and the distribution of void radii in order to compare the
relationship of these  statistics to sample distribution geometry.

The paper is organised as follows. We start the paper with a short
description of the basic formulae and methods we use in calculations
(Section 2). Then we use analytic models for the spectra and calculate
respective correlation functions to get a general picture of the
behaviour of the correlation function on large scales (Section 3).
Thereafter we describe geometrical models of the distribution of
clusters used to investigate properties of the correlation function
(Section 4). The discussion of the correlation function for these
models, the comparison of this function with the power spectrum and
void radius statistics, and the discussion of the influence of the
sample shape follow (Section 5).  Next we discuss errors of the
correlation function (Section 6).  Finally we discuss geometric
properties of the correlation function on small scales (Section 7).
The paper ends with the main conclusions.

\section{ BASIC FORMULAE}

The two-point correlation function $\xi(r)$ is defined as the 
excess over Poisson of the joint
probability of finding objects in two volume elements separated by $r$
and averaged over a very large volume (Peebles 1980). It is possible,
in practice, to determine this function only for a limited volume,
which is called the estimate of the correlation function.  We  shall
use the term ``correlation function'' for its estimate, and calculate
it using the classical formula:
$$
\xi(r)={\langle DD(r)\rangle  \over \langle RR(r)\rangle }{n_R^2 
\over n^2} -1, \eqno(1)
$$
where $\langle DD(r)\rangle $ is the number of pairs of galaxies (or
clusters of galaxies) in the range of distances $r\pm dr/2$, $dr$ is
the bin size, $\langle RR(r)\rangle $ is the respective number of pairs
in a Poisson sample of points,  $n$ and $n_R$ are the mean number
densities of clusters in respective samples, and brackets $\langle
\dots \rangle $ mean ensemble average. The summation is over the whole
volume under study, and it is assumed that the galaxy and Poisson
samples have identical shape,  volume and selection function.

We define our model samples in cubic volumes of side-length $L$, and
calculate the correlation function for the whole distance interval from
$r=0$ to $r=L$, dividing this interval linearly into 100 equal bins.
For a given sample shape and selection function we calculate $\langle
RR(r)\rangle /n_R^2$  as a mean value of ten realizations of random
samples of 2000 clusters with the same selection function as the model
sample. In this paper we use only simple selection functions: in
addition to cubical samples we consider double-conical samples (see
below).

To suppress random noise we  smooth the correlation function with a
Gaussian window.  After smoothing the Poisson error is virtually
absent and the residual error is determined by the cosmic variance of
the geometry of the particle distribution, i.e. by the variation
of the correlation function found for different volumes of space.

We calculate  also the power spectrum, $P(k)$, for our models. For this
purpose we can use the relation between the power spectrum and the
correlation function, as they form a pair of Fourier transformations:
$$
\xi(r)={1 \over 2\pi^2}\int_0^{\infty}{ P(k) k^2 {\sin kr \over kr} dk}, 
\eqno(2)
$$
$$
P(k)=4\pi\int_0^{\infty}{\xi(r) r^2 {\sin kr \over kr} dr}.
\eqno(3)
$$
Here the wavenumber $k$ is measured in units of $h~ {\rm Mpc^{-1}}$,
and is related to the wavelength, $\lambda=2\pi/k$. To find the power
spectrum from the correlation function we could use the Eq. (3).
However, it is more convenient to use direct methods which are similar
to the calculation of the spectrum of $N$-body models of structure
evolution, calculating first the distribution of clusters in the
Fourier space and then averaging over various wavenumber 
directions. The
distribution of clusters in $k$-space is found with the fast Fourier
transform using $64^3$ cells, which allows us to find 32 modes in the
spectrum. This method assumes that the whole space is filled with
similar periodic cubes.

In addition to geometric models where the distribution of clusters in
real space is given initially, we also find for comparison the relation
between the correlation function and power spectrum analytically for
some simple power spectrum models. In this case we use the Eq. (2). 

\begin{figure*}
\epsfysize=6.5 cm
\epsfbox [-10 60 620 320]{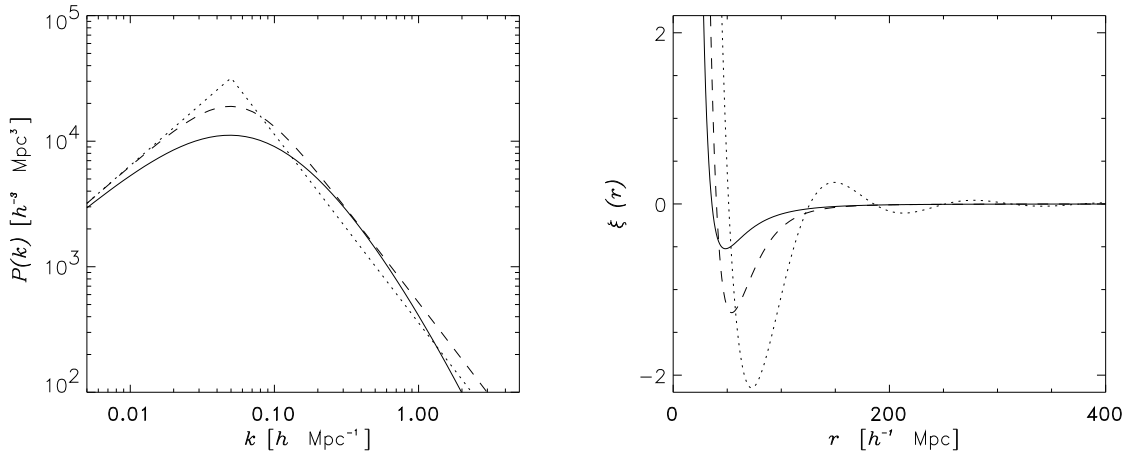}
\caption{
The power spectra (left) and correlation functions (right) for
a double power model with sharp transition (dotted line) and
with smooth transition (dashed line) and for the standard CDM model
 (solid line).  } 
\end{figure*}

\section{ ANALYTIC MODELS}

Following Eq. (2) we have calculated the correlation function for
several power spectra. In general, for any given power spectrum the
correlation function cannot be calculated analytically. However, we
can determine $\xi(r)$ using a simple ansatz for $P(k)$:
$$
P(k)=\cases{A k^n,&$k\le k_0$;\cr\cr
A k_0^n (k/k_0)^m,&$k>k_0$,} \eqno(4)
$$
where $n=1$ is the power index on large wavelengths
(Harrison-Zeldovich spectrum), the negative $m$ is the index on
small wavelengths, $k_0$ is the transition wavenumber, and $A$
is the normalisation amplitude.  From observations $m$ is
expected to be about $-1.5$. Thus, we have to integrate 
\setcounter{equation}{4}
\begin{eqnarray}
\xi(r) = & {A k_0^4 \over 2\pi^2} \left(  y^{-4} \int_0^y x^2 \sin x \,
 dx + \right.\nonumber\\
         & \left. y^{-3/2} \int_y^\infty x^{-1/2} \sin x \, dx \right),
\end{eqnarray}
where $x=kr$, and $y=k_0r$ depends on the place $k_0$ of the maximum of the
power spectrum.  The solution of the integral yields
\begin{eqnarray}
 \xi(r) = {A k_0^4 \over 2\pi^2} \bigl\{ y^{-4} &
          [ 2(\cos y + y \sin y -1) - y^{2} \cos y ] \bigr. \nonumber\\
          & \bigl. + y^{-3/2} [ \sqrt{\pi/2}
          - \sqrt{2\pi} \hbox{S}( \sqrt{2y/\pi}) ] \bigr\} ,
\end{eqnarray}
where S denotes Fresnel's sine-integral. Obviously, the
correlation function $\xi(r)$ is oscillating. The amplitude
scales with $k_0^4$ and the radial dependence with $k_0 r$. The
integrals in Eq.(5) (the power spectrum left and right of the
maximum) lead to approximately the same contribution in $\xi$,
however with opposite signs. The resulting oscillations are
rapidly damped ($\propto r^{-3}$).

As a second model we consider a smooth transition between
positive and negative spectral indices (Peacock \& West
1992, Einasto \& Gramann 1993)
$$
P(k)={A k^n \over 1+(k/k_0)^{n-m}}. \eqno(7)
$$
The third model is the standard cold dark matter model (SCDM)
with the transfer function of Bond and Efstathiou (1984), where
$\Gamma = \Omega h = 0.5$. The transition between positive and
negative spectral indices is slower than in the second model.
The maximum of the first two  models is chosen to be at $k_0 =
0.0562 h^{-1}$~Mpc in accordance with the SCDM model.  The
amplitude is taken in accordance with COBE data by adopting
$A=6.4\times 10^5\ h^{-4} {\rm Mpc}^4$ (White, Scott \& Silk
1994).

\begin{figure*}
\epsfysize=10.5 cm
\epsfbox[10 240 560 600]{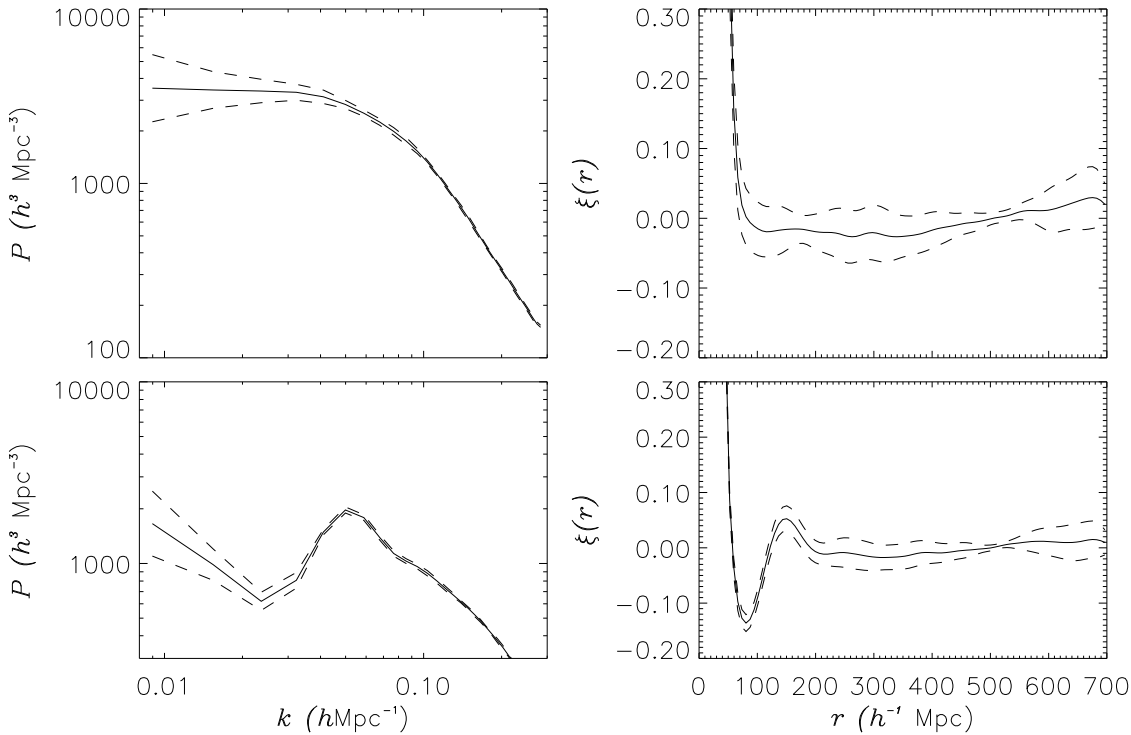}
\caption{
The power spectra (left panels) and correlation functions (right
panels) for the random supercluster model (upper panels) and the Voronoi
model (lower panels). Solid curves indicate the mean function for ten
realizations, dashed lines show the $1\sigma$ error corridor calculated 
from the scatter of individual functions.
}
\end{figure*}

A rapid and precise method to calculate numerically the integral (2) is
given by Press \etal (1992). The dispersion of density
fluctuations is equal to the value of the correlation function
at zero separation,
$$
\xi(0) = {1 \over 2\pi^2} \int_0^{\infty}{P(k) k^2 dk}. \eqno(8)
$$
In all cases of practical interest the spectral index is $m \ge -3$;
i.e. the integral (8) is not convergent. Since the dispersion of
density fluctuations has a finite value, we conclude that on small
scales the real spectral index must decrease so that the spectrum
becomes steeper. The physical reason for this behaviour is  galaxy
formation; i.e. the power spectrum arises from objects of finite size.
This effect can be formally described by an exponential cutoff of the
spectrum with an effective wavenumber $k_s=50$ $h$ Mpc$^{-1}$
corresponding to a scale $\approx 0.1$ \Mpc.  This parameter does not
change the correlation function on scales of interest for the present
study. The spectra and respective correlation functions are shown in
Figure~1.

A smooth transition between positive and negative spectral indices can
be approximated by a number of sections with constant spectral index
$n_i$. The index changes at wave-numbers $k_i$. Then the correlation
function is a superposition of given analytic  periodic functions with
different periods $2\pi/k_i$, so  the resulting correlation
function does not oscillate (Gottl\"ober 1996).

It is evident that the amplitude of the correlation function is
proportional to the amplitude of the power spectrum. Furthermore, in the
first example with an analytical solution, the scaling of the correlation 
function with $k_0$ is obvious. In general the power spectrum can be written 
as
$$
P(k) = A k T^2(\zeta k),\eqno(9)
$$
where $A$ is the normalisation constant, $\zeta$ defines the place of
the maximum ($\zeta = k_0^{-1}$ in the ansatz (4) and $\zeta \approx 20
(\Omega h^2)^{-1} \hbox{Mpc}$ for the CDM model, and $T$ is the transfer 
function. Thus the correlation function is a function of $x = r/ \zeta$,
$$
\xi(x) = \zeta^{-4} f(x),\eqno(10)
$$
where 
$$
f(x) = {1 \over x} \int_0^{\infty} T^2(y) \sin(yx) dy.  \eqno(11)
$$
All correlation functions scale in the same manner as the analytic
solution (6).

Let us finally consider a correction $P_1(k)$ to the power spectrum $P(k) =
P_0(k) + P_1(k)$ which leads to $\xi(r) = \xi_0(r) + \xi_1(r)$. Obviously, 
this correction leads to superimposed oscillations of the correlation 
function if it has the form of the power spectrum Eq. (4) (see Section
4.3). The addition of such a correction with a relative power of about
25 \% leads already to significant oscillations in the correlation
function.

\section{  GEOMETRICAL MODELS OF THE LARGE-SCALE STRUCTURE}

In this Section we describe and analyse the correlation functions and the
power spectra for a number of geometrical toy models of the distribution of
clusters of galaxies.  Models are geometrical in nature; i.e. in these models
not the power spectrum is given initially but the distribution of clusters.
Models can be divided into three basic types: random models, regular models,
and mixed models. In all models we explicitly generate clusters that in most
models are clustered into superclusters, and designate the models
correspondingly by capital letters CL or SC.  The following small letters in
the model designation indicate the character of distribution; there follow two
digits which indicate the dispersion of the location of superclusters around
the mean position. These designations shall be described in detail below.

For each model we generate ten random realisations, calculate the mean
smoothed correlation function, and estimate the parameters of the correlation
function as described below. We find also the local rms deviations of
individual correlation functions from the mean value.  These deviations define
the error corridor for the correlation function.

\begin{figure*}
\epsfysize = 10.5 cm
\epsfbox[10 240 560 600]{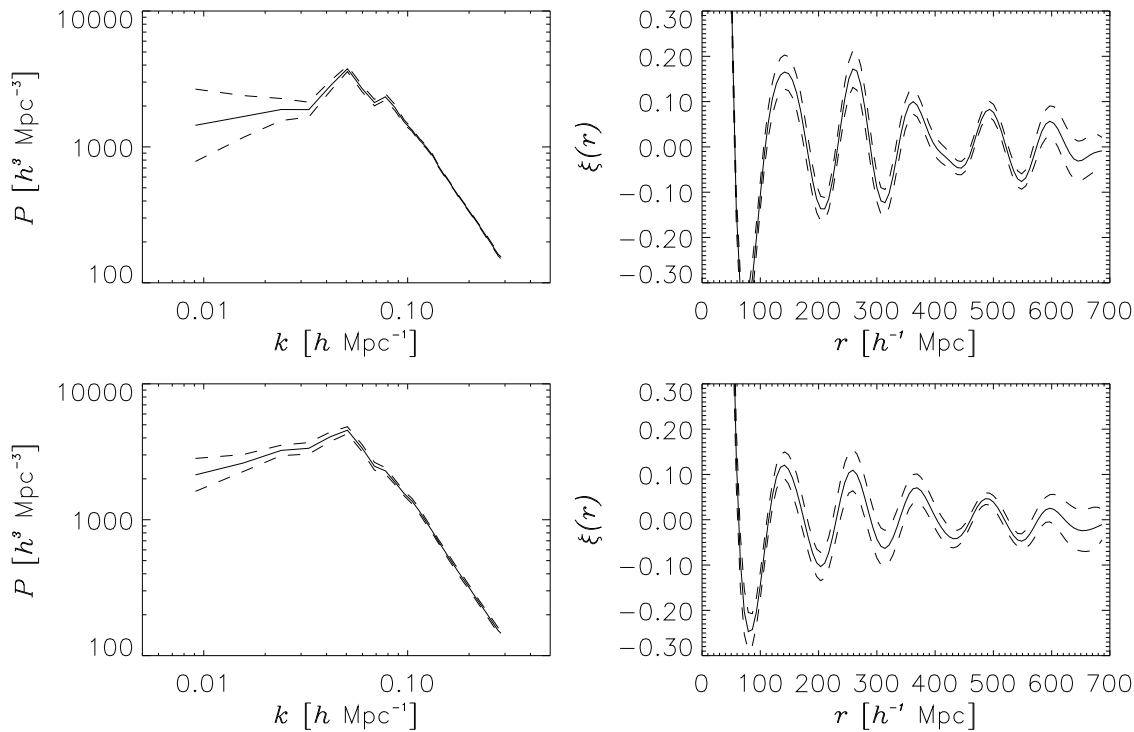}
\caption{
The power spectra and correlation functions for the regular net (upper panels)
and regular rod models (lower panels).
}
\end{figure*}

\subsection{ Random models}

Random models are designated by ``ran'' in the Tables and Figures. We
have generated random cluster, supercluster, and void models.

{\it  Random cluster} samples  are generated with the goal of checking
the formulae for the calculation of errors. The number of clusters was
taken to be from several hundred to several thousand. 

{\it Random superclusters.} In this model supercluster centres are positioned
randomly in rectangular coordinates.  The multiplicity of the supercluster
(the number of clusters in supercluster $N_{cl}$) was chosen randomly in
agreement with the multiplicity distribution of real superclusters, as derived
in Paper I.  The multiplicity distribution of real superclusters is from 1
(isolated clusters like the Virgo cluster) to 32 (the Shapley supercluster).
For simplicity only superclusters of richness $N_{cl} = 1, 2, 4, 8, 16, $ and
32 are generated. Member clusters of superclusters of richness 2 and higher
are located around the centre of the supercluster randomly with an isothermal
radial distribution.  Actual superclusters are elongated and we have checked
the influence of the elongation on the correlation function. In a model with
elongated superclusters negative sections of the correlation function have
smaller amplitudes. Positions of maxima and minima are not influenced. The
radius of superclusters was chosen as a slow function of the supercluster
richness, $R_{SC}=R_0 N_{cl}^{0.25}$.  The parameter $R_0$ determines the
cluster correlation radius; we used a value 15 \Mpc. These parameters
determine the cluster correlation function and the power spectrum on small
scales; values chosen yield good approximations to observed functions.  In a
cubic volume of size 700 \Mpc\ 1200 superclusters are generated, the total
number of clusters being a random number around 3000. In a sample of randomly
located superclusters some .superclusters are located closely together and
merge. To find the actual multiplicity distribution of superclusters a
clustering algorithm was applied, as for real cluster samples using a
neighbourhood radius 24 \Mpc. For details of supercluster definition see Paper
I.

{\it  Random voids} of diameter 100 \Mpc\ are distributed randomly and
clusters of galaxies are  located randomly outside voids, i.e. no
superclusters are generated. This diameter is equal to the mean
diameter of voids in samples of rich clusters of galaxies (Einasto
\etal 1994, Paper I). 

{\it The Voronoi tessellation} model is a variant of the random void
model. In this case centres of voids have also been located randomly,
but in contrast to the previous model clusters are not distributed
randomly outside voids. Instead, void centres are considered as seeds,
and clusters of galaxies are located in corners of a structure which
is formed by expanding volumes of seeds. Superclusters in this model
form by clustering of clusters if their mutual distances are smaller
than the neighbourhood radius 24 \Mpc\ used in the definition of
superclusters.  The Voronoi model has been investigated by a number of
authors (van de Weygaert \& Icke 1989; Ikeuchi \& Turner 1991;
Williams, Peacock \& Heavens 1991; SubbaRao \& Szalay 1992; van de
Weygaert 1994). We designate the model by ``vor''. It has one free
parameter, the number density of void centres. This determines the
mean separation of superclusters across voids which is taken as the
characteristic length of the model.  We generated ten models of box size
700 \Mpc\ with 431 void centres, in which case the characteristic
scale is 115 \Mpc, a value observed in our study of superclusters and
voids in Paper I, and in the study of the cluster correlation function
in Paper II.

Results for the correlation functions and power spectra of random
models are shown in Figure~2. All random models have a correlation
function with no sign of oscillations. The correlation function of the
random supercluster model approaches zero  on large scales
immediately after the initial maximum. The correlation function of the
Voronoi model is different. It has a minimum at a separation between
$\sim 60$ and $\sim 100$ \Mpc\ and one secondary maximum at $\sim 140$
\Mpc. We discuss differences between the random supercluster and
Voronoi models later.

The analysis of the correlation function of real cluster samples has
shown that it depends strongly on the richness of superclusters (see
 Paper II). We have thus made separate analyses of the correlation
function for our random models  for clusters located in rich and poor
superclusters. We shall discuss this problem in more detail below
(Section 7).

\subsection{ Regular models}

In these models clusters or superclusters are located, with a certain degree
of fuzziness, on the corners or on the edges between corners (rods) of a
regular rectangular three-dimensional grid. The fuzziness of
the model was realized by adding to positions of clusters or superclusters
random shifts around the exact position on the corner or the edge of the grid.
Depending on the degree of the fuzziness we can speak of strictly regular
models with no fuzziness or of quasi-regular models with variable fuzziness
(see also Einasto 1992).

{\it  Clusters on regular rods.}
In this model clusters of galaxies are located randomly along rods,
regularly spaced in a rectangular grid with a step $115$ \Mpc.  Rods
are randomly positioned around the exact values defined by the regular
grid with a scatter $\pm 20$ \Mpc, which is equal to the scatter of
mean diameters of voids defined by rich clusters of galaxies (Einasto
\etal 1994, Paper I). In our naming convention, the model is
designated by ``rod''.  Rods are cylindrical, and the diameter of
cylinders was taken equal to be the mean minor diameter of
superclusters, $\sim 20$ \Mpc\ (Einasto \etal 1994, Jaaniste \etal
1997).  Rods cross at the corners of the grid, and at these points
density enhancements form which can be considered as superclusters.

{\it  Superclusters on regular rods.}
In this model instead of clusters, superclusters are located randomly
along regularly spaced rods with step $115 \pm 20$ \Mpc.

{\it Superclusters in corners of a regular grid.}
In this  case superclusters  are located only in corners  of a  cubic
lattice with constant grid step. In our naming convention, this model
is designated as ``cor''.  Clusters of galaxies in superclusters are
generated as in the random supercluster model. Random shifts to
supercluster positions around corners are added with different values
of the mean shift.

{\it  Superclusters in a regular net.}
This model is a superposition of the model with superclusters located
in corners of the net, and additional random clusters located along
regular rods; this model is designated as ``net''. Clusters in rods
represent isolated clusters and cluster filaments which are
concentrated mostly between superclusters.  Here, too, variable random
shifts to mean positions in the regular net can be added.

The correlation functions and power spectra  for  regular models are
shown in Figure~3. 

The extreme case of regularity is a model with superclusters of equal
multiplicity  which are positioned exactly in corners of the regular
grid. The only free parameter it has is the step size of the grid.  
The correlation function and the
spectrum consist of a number of isolated peaks corresponding to mutual
distances of all possible pairs of distances of corners of the grid.
The first peak of the correlation function is located in a separation
equal to grid step, the next one to the shortest diagonal, and so on.
There exists no regular oscillations of the correlation function of the
type observed in analytical models with sharp maximum in the spectrum.
{\it This example shows that a strictly regular structure  cannot
produce  regular oscillations in the correlation function}.

\begin{figure*}
\epsfysize = 10.5 cm \epsfbox[10 240 560 600]{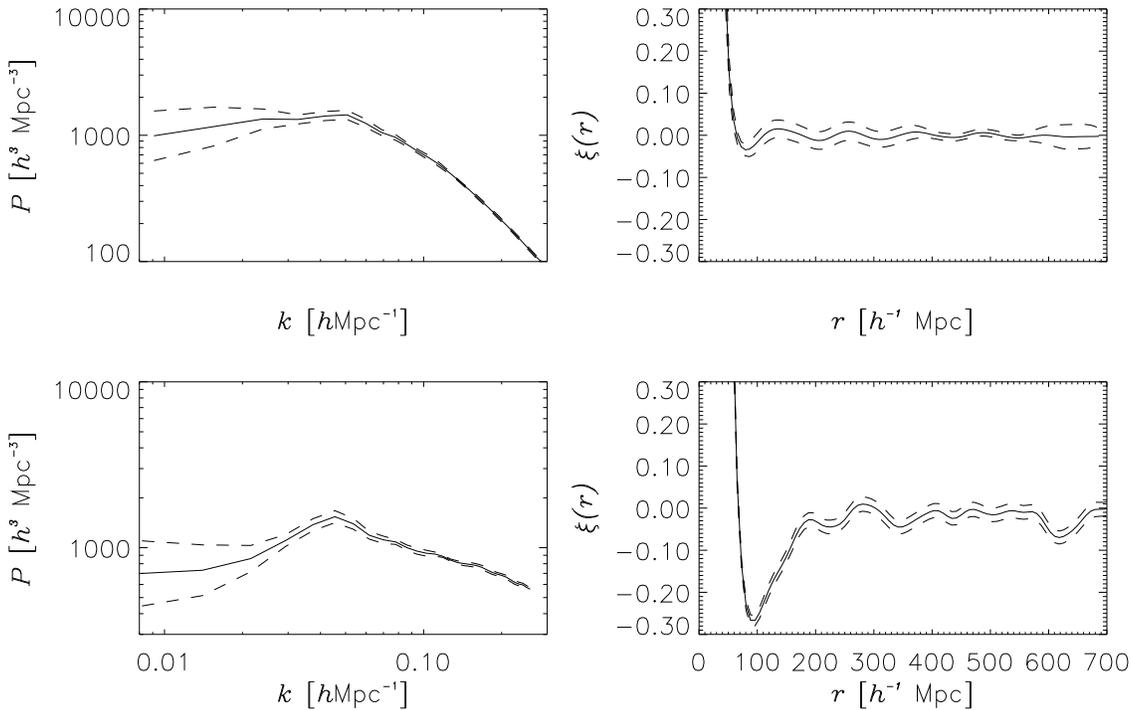}
\caption{
The power spectra (left panels) and correlation functions (right
panels) for the mixed model SC.mix.15 and the double power-law 
model DPS.6, shown in the upper and lower panels, respectively.
}
\end{figure*}

If we add some flexibility to the regular corner model by adding random
shifts to supercluster positions then neighbouring peaks of the
correlation function join to form broader maxima. Adding clusters on
rods makes the correlation function even more smooth. One such example
is shown in upper panels of Figure~3. If we make the geometry of
the model even more random and allow a random position of superclusters
on rods we get a correlation with even more regular oscillations, as
seen in the lower panels of Figure~3. This example shows that in order
to get a correlation function with regular oscillations there must be a
lot of freedom in the supercluster positions. These models are regular
only in a restricted sense (approximately equal step between rods).

With changing random shifts the amplitude of oscillations also changes.
It is  larger the more accurately superclusters are positioned in
corners or rods of the regular net. The scatter of superclusters around
the mean position used in our toy models is approximately equal to the
scatter of diameters of voids between real superclusters.  Thus our
models should have correlation functions with an amplitude of
oscillations which is close to the amplitude of oscillations of a
population of real clusters, if clusters form a regular network of
superclusters and voids with this scatter. The analysis made in Paper II
shows that this is indeed the case.

\subsection{  Mixed model}

The analysis of the correlation function of real cluster samples has
shown that clusters in high- and low-density environment have rather
different correlation functions. Clusters in high-density environment
(the members of very rich superclusters) have an oscillating
correlation function whereas clusters in low-density regions
(members of poor superclusters and isolated clusters) have an
almost zero correlation at large separations (Paper II).  To model this
behaviour of the correlation function we have constructed a mixed model.

The model has two populations of clusters: rich superclusters are located
randomly in regular rods and form the quasi-regular population, poor
superclusters are located randomly throughout the whole volume of the model.
For simplicity we use here randomly located superclusters as a low-density
cluster population, taking into account the fact that the correlation function
is not sensitive to details of the distribution of clusters in low density
environments (real clusters of the low density population lie near void walls,
and void interiors are empty of rich clusters, see Paper I). One third
of all clusters belong to the quasi-regular population, two thirds to the
random one.

The correlation function and spectrum for this model is presented in
Figure~4.  The correlation function is oscillating but with a very low
amplitude. The whole observed cluster sample has a similar correlation
function, and the amplitudes of oscillations are also similar (Paper
II).  This example shows that our mixed model is a good approximation
to the real cluster sample.

The mixed model is compared with a $N$-body model of structure evolution based
on the double power-law spectrum with a sharp transition between short and
long wavelength spectral indexes, Eq. (4). The model was calculated similarly
to the CDM model and was also used to estimate the cosmic variance of the
correlation function (see Section 6). The model is designated as DPS.6. The
oscillatory behaviour of the double power-law model is less regular; the first
minimum of the correlation function has much larger amplitude. Amplitudes of
maxima are approximately as high as in the case of the mixed model, the mean
period of oscillations is also similar, but the scatter of individual periods
is larger.

\begin{table*}
\begin{minipage}{120mm}
\caption{ Correlation function parameters of model samples}
\halign to \hsize {# \hfil &\hfil#\quad\hfil&\quad\hfil#\quad\hfil&
\quad\hfil#\quad\hfil&\quad\hfil#\quad\hfil&\quad#\quad\hfil&
\quad\hfil#\quad\hfil&\quad\hfil#\hfil&\quad\hfil#\hfil\cr
\noalign {\smallskip}
Model&$N$&Step&$r_{min}$&
$r_{max}$&$A_{max}$&$\Delta_{21}$&$\Delta_{32}$&$\Delta_{mean}$\cr
\noalign{\medskip}
\noalign{\smallskip}
CL.vor.00  &2920&115& 81&148&0.073& \cr
\noalign{\smallskip}
SC.cor.25  &1684&120& 64&146& 0.40&127&104&120\cr
\noalign{\smallskip}
CL.rod.20  &3240& 115& 82&138&0.185&119&109&116 \cr
\noalign{\smallskip}
SC.rod.15  &3003& 115 & 82&141&0.150&118&109&115 \cr
\noalign{\smallskip}
SC.net.00  &3452& 128 & 96&139&0.45&150&118&129\cr
SC.net.20  &2719& 115 & 74&142&0.175&119&101&116\cr
\noalign{\smallskip}
SC.mix.15  &4021& 115 & 82&136&0.022&121&115&116\cr
}
\end{minipage}
\end{table*}

\subsection { Parameters of oscillations of the correlation function}

We shall use the following parameters to characterise oscillations of
the correlation function: the mean separation of the first minimum from zero,
$r_{min}$; the position and amplitude of the first secondary maximum,
$r_{max}$ and $A_{max}$; differences between the subsequent maxima,
$\Delta_{ij}=r_{maxi} - r_{maxj}$; and the mean value of differences,
derived from the first 4 differences between maxima and from the first 4
differences between minima, $\Delta_{mean}$. Scaling parameters can
be used to derive the value of the true period of oscillations, $P$,
which we take to be equal to the grid step of the net of the
quasi-regular model, and to the wavelength of the maximum of the
spectrum, $\lambda_0=2\pi/k_0$.

Values of these parameters for models with at least one secondary minimum and
maximum are given in Table~1. They are derived from the mean smoothed
correlation function of the particular model, calculated from ten
realizations.  To avoid the decrease by smoothing the amplitude has been
estimated from unsmoothed data.

This Table shows that these parameters vary only within rather narrow
limits. The position of the first secondary maximum is almost the same
for all geometrical models with identical grid step (and period $P$).
Differences between maxima and minima are larger, but not much. These
differences are rather systematic: in all our geometric toy models
$\Delta_{21} > P$, and $\Delta_{32} < P$. The characteristic scale of
the supercluster-void network, $P$, can be calculated from the above
parameters using the mean values of the following parameters:
$$
f_1={r_{max} \over P} = 1.20 \pm 0.05; \eqno(12)
$$
$$
f_{21}={\Delta_{21} \over P} = 1.06 \pm 0.05; \eqno(13)
$$
$$
f_{32}={\Delta_{32} \over P}=0.93 \pm 0.05; \eqno(14)
$$ 
and 
$$f_m= {\Delta_{mean} \over P}=1.01 \pm 0.01. \eqno(15)
$$  
As we see the most accurate determination of $P$ comes from the
last equation.

\begin{figure*}
\epsfysize = 6.5 cm 
\epsfbox[10 240 560 430]{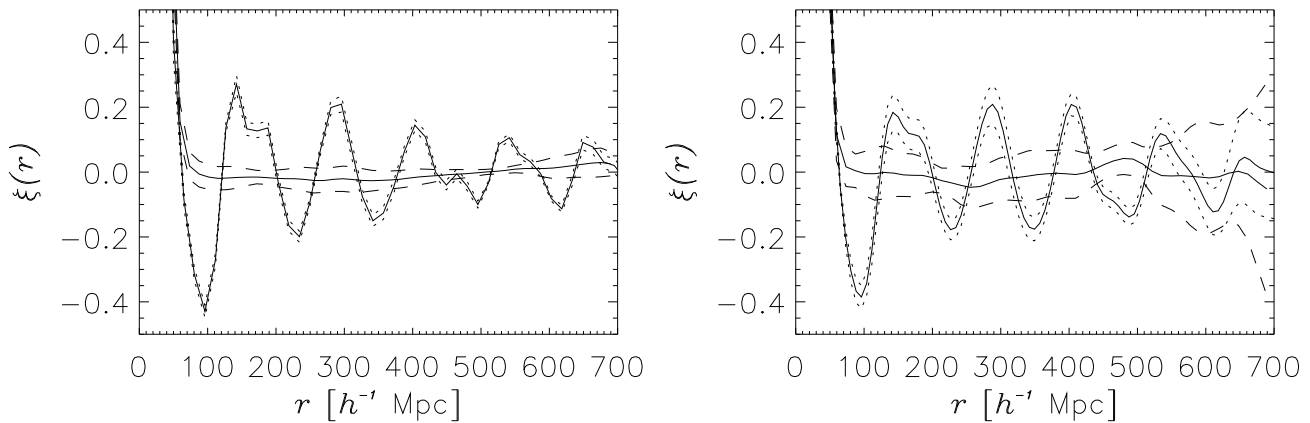}
\caption{
The correlation functions found for the whole cube (left panel) and
for  double conical samples (right panel). In  both panels these
functions are given for the regular net model and for
the random supercluster model. Error corridors for the correlation
function for the random supercluster model are plotted with 
long-dashed lines; error corridors for the regular net model 
with short dashed curves.
}
\end{figure*}

\section{ DISCUSSION}

In this Section we continue the analysis of the correlation function
and the power spectrum for different cluster distributions. Our main
goal is to find which geometric properties of the distribution of
superclusters can be detected on the basis of the correlation function
and the power spectrum alone. We investigate the influence of the shape
of the volume occupied with clusters. The available observed
samples are generally not cubical in shape, and the shape of sample
can, in principle, influence statistical properties of the samples.

\subsection{ The correlation function}

The correlation functions with their error corridors and the respective
power spectra with their error corridors are plotted for several models
in Figs~2 -- 4. On small scales all correlation functions have a large
maximum which reflects the fact that our models have clustering
properties. On large scales the correlation functions are different.
They have one of three principal forms: they become flat on large
scales directly from the maximum on small scales (random supercluster
model); have a minimum on intermediate scales followed by one secondary
maximum, and then smoothly become flat (Voronoi model); or obey an
oscillatory behaviour with alternating secondary maxima and minima of
decreasing amplitude (models with a  built-in regular structure of the type of
a rectangular lattice).

Geometric interpretation of the correlation function of the first type
is simple: here  superclusters are distributed randomly and peaks from
numbers of objects at various separations cancel each other out.

Superclusters separated by large voids produce a correlation function with a
minimum, whose location corresponds to the mean separation between
superclusters and voids. The minimum is followed by a secondary maximum
corresponding to the distance between superclusters across voids. On still
larger scales there are no regularities in the supercluster-void network
(voids are spaced randomly), and the correlation function on very large scales
is close to zero since superclusters at various separations cancel each other
out as in the previous case.  This distribution is realized in the Voronoi
model.

The interpretation of the oscillating correlation function is also
straightforward. In all models which produce an oscillating correlation
function the distribution of clusters on large scales is quasi-regular in the
sense that high-density regions form a fairly regular network with an
approximately constant grid length. In order to get regular oscillations the
geometric structure must not be too regular. An example of very regular
structure with superclusters located only in corners of the grid shows that
the correlation function has a number of small maxima and minima without any
regular oscillations.  Quasi-regular oscillations are generated in all models
with certain regularity and certain irregularity: the supercluster-void
network must have a constant overall scale but individual superclusters must
be located flexibly within this network. The scale of the model is determined
by the period of oscillations. To check the correspondence of the model to
reality one can use small deviations of the correlation function from strict
regularity expressed through parameters $\Delta_{21}$ and $\Delta_{32}$. As we
see (Paper II), deviations of the observed correlation function from strict
regularity are what is expected for models with constant scale and a certain
scatter in individual superclusters positions. This coincidence cannot be
accidental, since a visual inspection of the distribution of real
superclusters also shows the presence of such a network of superclusters and
voids (Paper I).

The basic difference between the regular and Voronoi models lies in
the correlation of positions of high-density regions on large scales:
in the regular model positions are correlated and in the Voronoi model
not.

\subsection{ Influence of the shape of the sample}

Real cluster samples are limited to a certain maximal distance, and
clusters in the galactic equatorial zone are not seen due to
obscuration. Thus real observational samples generally have a double
conical shape. To investigate the influence of such a sample shape we
selected from cubical cluster samples double conical subsamples.  These
subsamples have a conical height equal to  half the size of the total
sample, and the equatorial belt (galactic latitudes less than $\pm
30^{\circ}$) is excluded. The volume of these subsamples is about
one-quarter the volume of the whole cube, and they contain about
one-quarter the number of clusters than is in the original sample.

In Figure~5 we compare correlation functions of two models -- the regular net
model and the random supercluster model -- calculated for the whole cubical
volume, and for a double conical volume. The first model is regular and the
corresponding correlation function is oscillating; the second model is not.
This Figure shows that the oscillating properties of the sample are preserved
in the smaller double conical sample. Only the error corridor is larger in the
smaller samples, as expected. This similarity of the oscillating properties is
due to the fact that the small samples are large enough to contain in
sufficient quantity the structural elements responsible for oscillations.

\begin{figure}
\epsfysize = 5.5 cm 
\epsfbox[70 290 570 620]{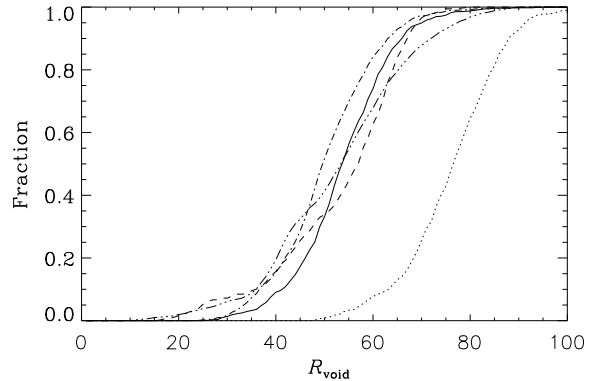}
\caption{
The integral distribution of void radii for selected models.
Models are plotted as follows: solid line -- Voronoi model CL.vor.00;
dash-dotted line -- random supercluster model SC.ran.20; dashed
line -- superclusters on rods model SC.rod.20; dash-dot-dot line --
superclusters in net SC.net.20; dotted line -- superclusters in corners
model SC.cor.25.
}
\end{figure}

\subsection{ The power spectrum}

Spectra for models are plotted in Figures~2 -- 4.  Power spectra have three
regions, corresponding to inhomogeneities with high, medium, and low spatial
frequency (or small, medium, and long wavelengths). This division is given by
the maximum of the spectrum.  Our calculations show that on small wavelengths
(large wavenumber $k$) spectra are rather similar. This result is expected as
perturbations on these wavelengths are determined by the internal distribution
of clusters within superclusters, and superclusters in our models are
generated in a similar way.

Calculations by Frisch \etal (1995) show that inhomogeneities on very
large wavelengths do not influence the distribution of clusters in
models, instead superclusters are modulated, i.e. in some regions they are
very massive and in others poor.  Regularity of the structure is given
by power on intermediate wavelength. 

Analytic calculations confirm this result that models with only slightly
different power spectra in the medium wavelength region may have very
different correlation functions. The mixed model with quasi-regular network of
rich superclusters and voids has a spectrum which differs from the spectrum of
the random supercluster model only by about 25 \% near the maximum.
Differences between power spectra of the Voronoi and regular rod models are
also small whereas correlation functions are very different.  Such small
differences can result from different geometries of the supercluster
population.  These examples show that the correlation function is much more
sensitive to the presence of a small regularity in the distribution of
clusters than it is to the spectrum itself.

\subsection { The distribution of voids}

One feature of the correlation function is not yet explained. Why does the
Voronoi model show a minimum and maximum in the correlation function and
the random supercluster model does not? One possible reason could be a
difference in the distribution of voids. In the random supercluster model
voids can be very small, which is not the case for the Voronoi model. To
check this possibility we extracted voids from our model cluster samples
and calculated the integrated distribution of void radii. Here we used
procedures described  by Einasto, Einasto \& Gramann (1989) and
Einasto \etal (1991). Results for the void radii are shown in Figure~6.

Our calculations show that for most of our models the distribution of
void radii is very similar. A factor which influences the distribution
of void radii is the presence of weak filaments between superclusters.
If clusters in high-density regions are distributed similarly in
different models then their correlation functions also are similar, as they
are, for instance, for models SC.cor.25 and SC.net.20. Void diameters
in these models are, however, completely different, as seen in
Figure~6. In  models with no filaments between superclusters
(SC.cor.25) voids are much larger -- their mean diameter corresponds to
the diagonal of the net -- whereas in all models where rods between
corners are also populated, void diameters are determined by distances
between these rods, i.e. by the scale of the net.

We conclude that voids are determined by properties of small filaments and the
correlation function by the distribution of clusters in high-density regions
(see also Paper I).

\begin{figure*}
\vskip -2mm
{\epsfysize = 6.5 cm \epsfbox[20 250 570 440]{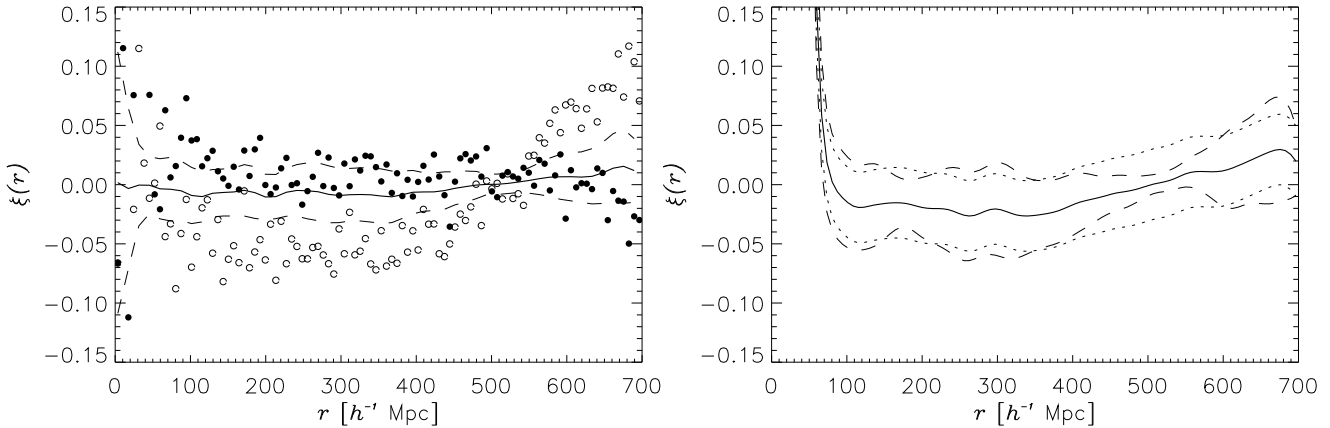}}
\caption{
Random and cosmic errors of the correlation function of model samples.
In left panel  we plot by open and filled circles values of the
unsmoothed correlation function for two realizations of random samples
of 1000 clusters. The solid curve is the mean smoothed correlation
function for ten realizations using a Gaussian kernel with dispersion
$\sigma=15$ \Mpc. Dashed lines around the smoothed curve show the error
corridor calculated from the scatter of ten realizations.  In the right
panel we show the mean smoothed correlation function for the random
supercluster model; dotted lines show the error corridor
calculated from the Eq. (16), and dashed curves are for the error
corridor calculated from the scatter of individual correlation
functions for ten realization.
}
\end{figure*}

\section{ ERROR ESTIMATES}

The errors in the two- and three-point correlations are discussed by
Mo, Jing \& B\"orner (1992). Errors in the correlation function depend
on high-order correlations, which are different for different geometrical
distributions of clusters. The difficulty of the determination of errors 
lies in the fact that they depend on the distribution of clusters in the
ensemble, and there is no unique comparison ensemble of samples
available. Thus parameters describing the errors of the correlation
function must be determined for all different geometrical distributions
of clusters.

In the calculation of the correlation function the number of
particles in the  Poisson sample is taken as very large. In this case we
can ignore errors of the Poisson sample and calculate the error of the
correlation function  from the error of the number of pairs $\langle
DD(r)\rangle $.  Mo \etal give a formula to calculate the error of $\langle
DD(r)\rangle $ through moments of the two-point and three-point
correlation function. The first terms have the form:
$$
\sigma_{DD}^2(r)= \langle DD(r)\rangle +{b^2\over N}
\langle DD(r)\rangle ^2. \eqno(16)
$$
Here $N$ is the total number of clusters in the sample, and $b$
is a constant which depends on the high-order correlation functions.
The first term of the Eq. (16) is the Poisson error, due to random
errors in sampling of galaxies; the second term is the cosmic error or
variance, due to variations in the distribution of clusters in
different parts of the Universe.

For our models we have calculated  errors in the correlation function
from the scatter of different realizations of a particular model, and
the Eq. (16) was used only as a convenient parametrisation to
determine the value of the parameter $b$. As we see below this formula
is quite accurate and gives results in good agreement with direct
estimates. To see better the dependence of errors on the number of
clusters in the sample we find approximate formulae for random and
cosmic errors of the correlation function on large scales. On these
scales the correlation function $\xi(r)\approx 0$, and thus
$\langle RR(r)\rangle (n^2/n_R^2) \approx \langle DD(r)\rangle $.  We
get for the random error in the correlation function:
$$
\sigma_{\xi P}(r)\approx {1 \over \sqrt{\langle DD(r)\rangle}}, \eqno(17)
$$
and for the cosmic variance:
$$
\sigma_{\xi c}\approx {b\over \sqrt{N}}. \eqno(18)
$$

The number of object pairs at separation $r$ is proportional to the
number of clusters squared, $\langle DD(r)\rangle \sim N^2$,  thus
$\sigma_{\xi P}\sim  N^{-1}$.  Random errors depend on the bin length
$dr$ and on the separation $r$. To reduce random errors we have smoothed
the correlation function with a Gaussian kernel of dispersion $\sigma$,
which can be varied within certain limits.

\begin{table}
\caption{ Cosmic variance parameter}
\halign to \hsize {# \hfil &\hfil#\quad\hfil&\quad\hfil#\quad\hfil&
\quad\hfil#\hfil\cr
\noalign {\smallskip}
Model&$\langle N\rangle $&$\sigma_{\xi c}$&$b$\cr
\noalign{\medskip}
CL.ran.00         &1000& 0.014&0.64 \cr
\noalign{\smallskip}
SC.ran.00         &2890& 0.029&1.58 \cr
\noalign{\smallskip}
CL.vor.00         &2920& 0.020&1.08\cr
\noalign{\smallskip}
SC.cor.25         &1684& 0.040&1.62\cr
\noalign{\smallskip}
CL.rod.20         &3240& 0.009&0.53\cr
\noalign{\smallskip}
SC.rod.15         &3003& 0.032&1.75\cr
\noalign{\smallskip}
SC.net.00         &3452& 0.017&1.00\cr
SC.net.20         &2719& 0.029&1.54\cr
\noalign{\smallskip}
SC.mix.15         &4021& 0.017&1.08\cr
\noalign{\smallskip}
DPS.6              &2388& 0.020&0.98\cr
\noalign{\smallskip}
CDM.61           &2588&0.036&1.83\cr
}
\end{table}

The cosmic variance is independent of the bin size and scale. It
depends only on the geometry of the distribution of systems of clusters
in the sample. The constant $b$ increases with the size of the
irregularities in this distribution.

The influence of random and cosmic errors is illustrated in Figure~7.
In the left panel we compare the effect of random and cosmic errors for
samples of random clusters. The scatter of individual values in the
unsmoothed correlation function (circles) shows the effect of random
errors, the overall deviation  of open and filled circles from the mean
correlation function (solid curve) is a visual demonstration of the
effect of the cosmic variance.

The influence of the smoothing length is studied in more detail in the
accompanying paper devoted to the analysis of real cluster samples (Section
4.4 in Paper II).  The number of clusters in observed samples is in some cases
only several hundreds, thus random errors are larger. For observed samples a
smoothing length about 15 \Mpc\ reduces the random noise in practice but does
not influence parameters of the correlation function.  Our models are designed
for comparison with observations, thus we use a smoothing length 15 \Mpc\ in
most model samples too.

In the right panel of Figure~7 we compare the error corridors found from the
scatter of individual realizations of the random supercluster model and
calculated from the Eq. (16) or (18) (both formula give practically identical
results on large scales).  We see that the widths of both error corridors are
approximately the same.  This shows that the Eq. (18) is a good approximation
of the cosmic variance.

To derive the possible cosmic variance we have used not only geometrical toy
models but also several $N$-body simulations of structure formation and
evolution. These simulations were performed with $128^3$ particles and $256^3$
grid cells using a particle-mesh code.  The size of the computational cube was
taken to be $L=700$~\Mpc\ $ \approx 6\lambda_0$, where $\lambda_0$ is the
characteristic scale of the supercluster-void network.  Clusters of galaxies
were selected by a procedure similar to the ``friends-of-friends'' algorithm.
The lower limit of the cluster masses in the simulations is determined by the
number of clusters in the sample. The cube was divided into 8 subcubes of half
the size of the original cube; perturbations within subcubes can be considered
as independent of each other. The correlation function of clusters of galaxies
was calculated for all subcubes, and the cosmic variance was found from the
scatter of results of individual subcubes. The power spectrum for the sample
of clusters of galaxies was calculated as for all other cluster samples
separately for all subcubes, and then the mean value and its scatter were
found.

In simulations we used a  CDM spectrum with the standard transfer
function (Bond \& Efstathiou 1984). This model is  COBE normalized in
amplitude at long wavelengths, and it has one free parameter,
$\Gamma=\Omega h$, depending on the total density of the model
$\Omega$ and the Hubble constant $h$. The model is designated as
``CDM.61'', where 6 designates the size of the computational box in
units of the scale of the supercluster-void network, and  1 is for the
standard high-density model ($\Omega=1$).

To estimate the value of the parameter $b$ from the cosmic variance
Eq. (16)  we found the mean $1\sigma$ rms deviations of individual
values from different realizations of the correlation function for the
scale range $r/L=0.25$ to $r/L=0.85$. (Here $L$ is the size of the
computational box.)  In this scale range  rms deviations are
approximately constant. The parameter $b$ is  given in Table~2.

The parameter $b$ depends primarily on the regularity of the
distribution on large scales. If there are no superclusters present
(random cluster and clusters in rods models), the parameter is less
than 1. In the Voronoi model and regular net model without shifts the
error parameter has a value $b\approx 1$. In models with superclusters
which have more freedom in their position (say, to move along rods)
the parameter has a larger value $b\approx 1.8$.  The CDM.61 model
also has a value of the parameter $b$ close to this.  In this case,
however, the amplitude of oscillations is smaller than for the sample
of clusters in high-density regions. A net model (superclusters in
corners with additional clusters on rods) with scatter 15 \Mpc\ around
the mean position is probably closest to the real case.

To determine from a sample of objects the presence of oscillations of
the correlation function, the cosmic error must be smaller than the
amplitude of oscillations.  If we adopt $b=1.6$ and assume that
oscillations have an amplitude 0.28 (Paper II), then the sample must have
at least 30 clusters. If we have a mixed population with an expected
amplitude of oscillation 0.05, then such oscillations can be
detected with confidence if the sample has at least $1000$ clusters.
We see that to detect oscillating properties it is useful first to
isolate the population with highest level of oscillating nature.

\section{ PROPERTIES OF THE CORRELATION FUNCTION ON SMALL SCALES}

Our main emphasis was the study of properties of the correlation
function on large scales. Our models contain, however, information also
on the correlation function on small scales. Thus it is of interest to
have a look on small scales, too.

As it is well known, on small scales the correlation function can be
approximated by a power law.  This law can be interpreted in terms of the
fractal distribution of galaxies. The fractal dimension is related to the
power index (Szalay and Schramm 1985). The fractal description is the
principal geometric interpretation of the correlation function on small
scales, and describes the shape of systems of galaxies. On small scales
spheroidal systems of galaxies dominate, whereas on larger scales the main
structural elements are filaments of galaxies (Einasto 1992).

It is well known that clusters of galaxies
have much larger value of $r_0$ than galaxies (Bahcall \& Soneira 1983,
Klypin \& Kopylov 1983). This result is usually interpreted as an
indication of the  {\it stronger clustering} of clusters of
galaxies.   However, Einasto \etal (1994) and Jaaniste \etal (1997)
showed that the mean minor diameter of superclusters is 20 \Mpc. 
We have calculated the correlation length separately for clusters
located in rich and poor superclusters, both in geometrical models and CDM
simulations. As in the real case, supercluster richness was determined by
the clustering analysis. Our calculations show that the correlation
length is very different for these populations, about 17 \Mpc\ for
clusters in poor superclusters and about 45 \Mpc\ for clusters in rich
superclusters. Similar results come from the analysis of real clusters
in Paper II.

Thus, a correct geometric interpretation of the correlation length of clusters
of galaxies is to say that systems of clusters of galaxies (superclusters) are
{\it larger} than systems of galaxies (clusters) since the parameter $r_0$
depends primarily on the size of the respective systems.

\section{ CONCLUSIONS}

Our study  shows that the correlation function has the following basic
properties.

 On large scales the correlation function of clusters of galaxies
depends on the geometry of the distribution of superclusters.  If
superclusters are located in a quasiregular net  (supercluster-void
network) then the correlation function has an oscillatory behaviour.
The period of oscillations is equal to the length of the step size of
the net. The amplitude of oscillations depends on the scatter of
distances between superclusters in the net and on the presence of a
more smoothly distributed population of  clusters.

 Parameters of oscillation and the scale of the net can be determined
from observations if the number of clusters is sufficient to get an
error corridor of the correlation function which is smaller than the
amplitude of oscillations. For a  quasiregular population this minimal
number of objects is about 30, and for a heterogeneous population it is
about 1000, depending on the fraction of the quasiregular component in
the whole population of clusters.

If superclusters are located in the corners of a random cellular void
network (Voronoi model) then the correlation function has a minimum, and its
location corresponds to the mean separation between superclusters and voids.
There follows a secondary maximum corresponding to the clustering of clusters
on opposite sides of voids.

If the distribution of superclusters is random then the correlation
function is featureless and becomes flat after the initial highly
positive value.

Oscillating properties of the correlation function are related to
the shape of the power spectrum near the maximum of the spectrum.
Oscillations occur only in the case when the spectrum has a
sharp maximum and sudden transition of the spectral index from
a positive value on large wavelengths and a negative one on short
wavelengths.

Thus we see that the correlation function is indeed a useful statistic
of the geometry of large scale structures. Observationally, we have
employed the large scale correlation function of a sample of Abell
clusters (E97, Paper II); we have exploited these results in a study of
the correlation function of Las Campanas Redshift Survey galaxies
(Tucker \etal 1995, 1997).

\section*{Acknowledgments}

This work was supported by Estonian Science Foundation grant 182 and
International Science Foundation grant LLF100.  JE and AS were supported in
Potsdam by the Deutsche Forschungsgemeinschaft; AS was supported by
the Russian Foundation for Basic Research  Grant 96-02-17591.  We thank
R. van de Weygaert for his Voronoi tessellation model program, and K.-H.
B\"ohning for his help to prepare figures.

\end{document}